\documentclass[prl,reprint,
 superscriptaddress, twocolumn,
aps,showpacs]{revtex4-1} 

\usepackage{graphicx}
\usepackage{amsmath}
\usepackage{url}
\usepackage{color}
\usepackage{verbatim}
\usepackage{gensymb}
\usepackage[utf8]{inputenc}
\usepackage{multirow}
\usepackage[english]{babel}
\usepackage{amsfonts}
\usepackage{amssymb}
\usepackage{enumerate}
\usepackage{sidecap}
\usepackage{bbold}
\usepackage[normalem]{ulem}
\usepackage{mathtools}
\usepackage[usenames, dvipsnames]{xcolor}

\def\beq {\begin{equation}}
\def\eeq {\end{equation}}
\def\beqar {\begin{eqnarray}}
\def\eeqar {\end{eqnarray}}
\def\w {\omega}

\def\bfq {\mathbf{q}}

\def\bfr {\mathbf{r}}

\date{\today}
\begin{document}

\title{Short-range excitonic phenomena in low-density metals} 

\newcommand{\lsi}{LSI, CNRS, CEA/DRF/IRAMIS, \'Ecole Polytechnique, Institut Polytechnique de Paris, F-91120 Palaiseau, France}
\newcommand{\etsf}{European Theoretical Spectroscopy Facility (ETSF)}
\newcommand{\soleil}{Synchrotron SOLEIL, L'Orme des Merisiers, Saint-Aubin, BP 48, F-91192 Gif-sur-Yvette, France}
\newcommand{\linz}{Institute for Theoretical Physics, Johannes Kepler University Linz, Austria}

\author{Jaakko Koskelo}
\affiliation{\lsi}
\affiliation{\etsf}

\author{Lucia Reining}
\affiliation{\lsi}
\affiliation{\etsf}

\author{Matteo Gatti}
\affiliation{\lsi}
\affiliation{\etsf}
\affiliation{\soleil}
 
\begin{abstract}
Excitonic effects in metals are commonly supposed to be weak, because the Coulomb interaction is strongly screened. We investigate the low-density regime of the homogeneous electron gas, where low-energy collective excitations and  ghost modes were anticipated. Using the Bethe-Salpeter equation (BSE), we show that both phenomena exist thanks to reduced screening at short distances. This is not captured by common approximations used in \textit{ab initio} BSE calculations, but requires vertex corrections that take the fermionic nature of charges into account. The electron-hole wavefunction of the low-energy modes shows strong and very anisotropic electron-hole correlation, which speaks for an excitonic character of these modes. The fact that short-range physics is at the origin of these phenomena explains why, on the other hand, also the simple adiabatic local density approximation to time-dependent density functional theory can capture these effects.
\end{abstract}

\maketitle
Electronic excitations in semiconductors and insulators often lead to excitons, electron-hole pairs that are bound by the  
screened Coulomb interaction and that dominate the low-energy range of absorption spectra in 
 insulators or low-dimensional materials, where the dielectric  screening is weak \cite{knox1963a,bassani1975a,Bechstedt2014}.
Excitons play a crucial role as energy carriers in many optoelectronic devices, such as solar cells, light-emitting diodes, single-photon emitters, etc.
In metals, instead, which behave approximately like a homogeneous electron gas (HEG), one would expect that excitons cannot form  because the macroscopic screening by free carriers is perfect \cite{Mahan1981,fetter2003a,giuliani2005a}, which suppresses the long-range electron-hole attraction \cite{nozieres1999a}. The dominant 
excitations of the HEG are instead plasmons, collective oscillations at higher frequencies, which originate from coupling of density changes by the long-range bare Coulomb interaction \cite{Pines1952,Pines1963}.

However, theory predicts a more complex and challenging phase diagram for the HEG \cite{ceperley1980a,ortiz1999a,Zong2002,Drummond2004,Holzmann2020,giuliani2005a,Azadi2022}. In particular, 
at low densities where the Wigner–Seitz radius $r_s > 5.25$, the static dielectric function $\epsilon(q,\w=0)$ for non-vanishing wavevector $q$ becomes negative \cite{giuliani2005a,dolgov1981a,Ichimaru1982,Takada2005,takada2016a}. This is not a classical phenomenon, and it can therefore never happen in the Random-Phase Approximation (RPA). It is crucial for the physics of the HEG: 
this ``dielectric catastrophe'' regime is the precursor of ground-state instabilities \cite{dolgov1981a,Ichimaru1982,Takada2005,takada2016a}, leading to symmetry breakings such as charge and spin density waves and Wigner crystallisation \cite{Wigner1934,Overhauser1968}. 
Since a strong enhancement of the static density and spin response functions induces those electronic instabilities, the investigation of the static dielectric response delivers crucial information about possible phase transitions \cite{Thouless1972,Sawada1961,giuliani2005a,Shore1978,Sander1980,Iyetomi1981,Ichimaru1982,Perdew1980,Perdew2021}.

The negative screening occurs when the dielectric function develops imaginary poles  that were originally called ghost plasmons \cite{dolgov1981a,takayanagi1997a}.
 More recently, Takada \cite{Takada2005,takada2016a} assigned these poles to excitations of excitonic, rather than plasmon-like, character. Their fingerprint should be a collective mode peaking in the negative imaginary part of the inverse dielectric function  $-{\rm Im}\,\epsilon^{-1}(q,\w)$ at small energies $\w$ and large wavevectors $q$, which might be detected experimentally by inelastic X-ray scattering  or electron energy loss spectroscopy \cite{schuelke2007}. The existence of excitonic modes would contradict the intuitive argument of the electron-hole interaction being completely screened out in the HEG. Indications for exciton formation are given by experiments, in particular,
time-resolved experiments that have identified transient excitons at metal surfaces \cite{Cui2014}.  Still, the impact of excitonic effects for valence electrons of metals remains largely unexplored, also theoretically \footnote{For core electrons, instead, see e.g. the classic papers by Mahan, Nozi\`eres and De Dominicis \cite{mahan1967a,nozieres1969a}.}. 

Takada's predictions are based on a formulation that is similar to time-dependent density functional theory (TDDFT), where a starting independent-particle response function is used in a Dyson-like screening equation with an effective interaction kernel $v_c(q) + f_{\rm xc}(q,\w)$ consisting of  the bare Coulomb interaction $v_c$, and the exchange-correlation (xc) contribution $f_{\rm xc}$ \cite{Petersilka1996} for which an approximate parametrized form is used 
\footnote{To be precise, following Ref. \cite{richardson1994a}, Takada dresses the independent particles through modified occupation numbers and adapts $f_{\rm xc}$ accordingly, with an improved version of the parametrization given in \cite{richardson1994a}.}.  
Such an approach should yield
spectra in principle correctly and in practice of good quality, according to the advanced level of approximation, but the compact equations and the physics hidden in the parametrization make it difficult to
analyze the spectra, confirm or infirm the excitonic character of collective modes, and identify the mechanism of exciton formation. Therefore,  the present work will address three challenging questions:
 {\it What justifies to associate the ghost excitation to an exciton? How can such a mode develop in a metal, i.e., which are the key ingredients that produce it? And therefore, how can we predict it, in which situation could we hope to find it, and what might be other measurable consequences and maybe potential applications?}

\begin{figure}[ht]
\begin{center}
\includegraphics[angle=270,width=0.98\columnwidth]{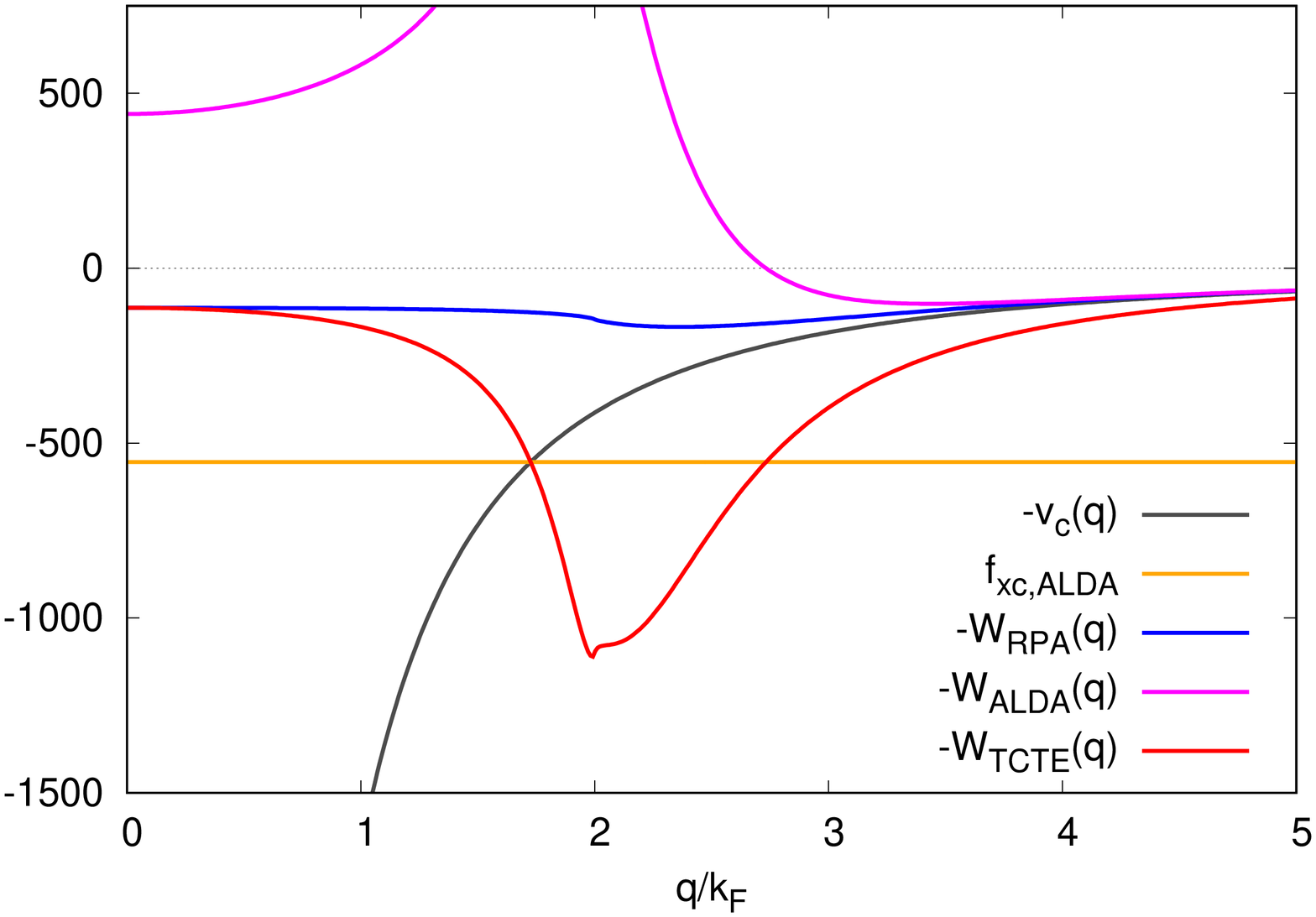}\\
\includegraphics[angle=270,width=0.98\columnwidth]{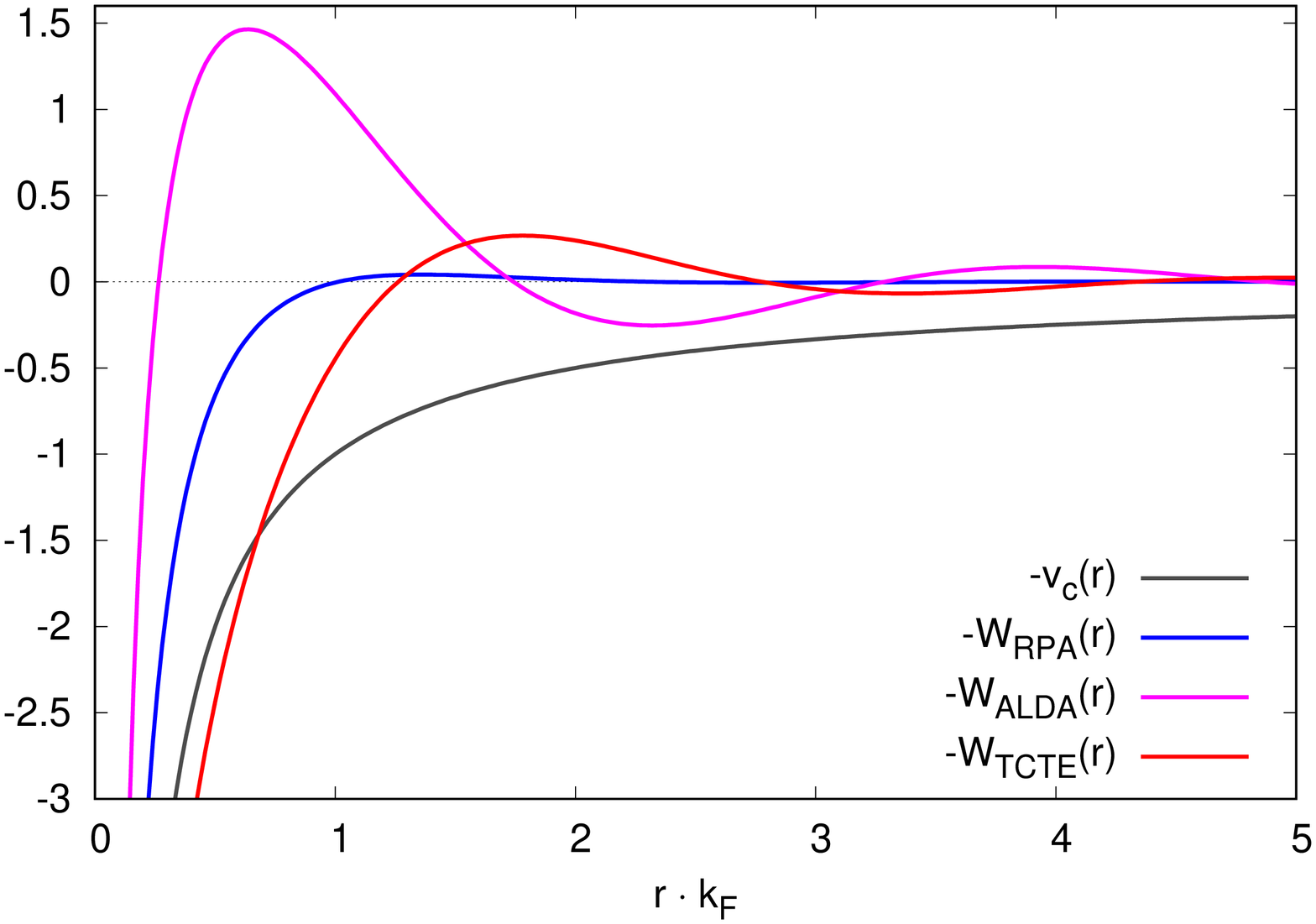}
\caption{Direct electron-hole interaction $-W^{\rm in}$ used in the BSE for the HEG at $r_s=22$, in reciprocal space (upper panel) and in real space (bottom). In grey $-v_c$, the bare Coulomb interaction. $-W^{\rm in}$ is shown in RPA (blue), ALDA (magenta), and TCTE calculated using the ALDA kernel (red).
The horizontal orange line in the upper panel is $f_{\rm xc}(q\to 0,\omega=0)$ in the ALDA. 
}
\label{fig:W}
\end{center}
\end{figure}

The main first principles formalism to describe excitons in materials is the Bethe-Salpeter equation (BSE) \cite{martin2016a,Bechstedt2014}, where the propagation of electrons and holes is governed by a self-energy $\Sigma$, and their effective interaction stems from the variation of $\Sigma$ with respect to the one-body Green's function $G$. Most often, $\Sigma$ is used  in the GW approximation \cite{hedin1965a,strinati1988a}, where the effective electron-electron (e-e) repulsion and electron-hole attraction (e-h) are given by the screened Coulomb interaction $\pm W(q,\w)= \pm \epsilon^{-1}(q,\w)v_c(q)$. Additionally, a quasiparticle approximation is made for electrons and holes, and the e-h attraction is taken at $\w=0$. Since $\epsilon$ is also the output of the BSE, a typical GW-BSE calculation follows the upper option in scheme \eqref{eq:bse-scheme}: 
\begin{align}
\begin{rcases}
    \epsilon^{\rm in}(q,\w) \to \Sigma=iGW^{\rm in} : \,\,\, \textrm{e-e  repulsion} \\
    \epsilon^{\rm in}(q,\w=0) \to -W^{\rm in} \, : \,  \,\,  \textrm{e-h attraction}
   \end{rcases} 
   \xrightarrow[]{\textrm{BSE}}  \epsilon^{\rm out}(q,\w) \nonumber \\
    \textrm{Alternative: }\,\, f_{\rm xc}(q,\omega) \xrightarrow[\hspace*{3.6cm}]{\textrm{TDDFT}}  \epsilon^{\rm out}(q,\w) 
    \label{eq:bse-scheme}
\end{align}
Often, an approximation such as the RPA is used for the input screening $\epsilon^{\rm in}$.
The intuition that excitons do not exist in metals is based on the Thomas-Fermi picture of perfect macroscopic static screening of the HEG \cite{Mahan1981,fetter2003a,giuliani2005a}, so $1/\epsilon^{\rm in}(q\to 0,\w=0)=0$. However, little definite knowledge exists, as few BSE calculations in metals can be found besides work on optical properties \cite{marini2003a,Uimonen2015}, including low dimensional materials \cite{Deslippe2007,Liang2014}, or work on the correlation energy \cite{Maggio2016}. Importantly, screening is complete only at very large distance. Indeed, $1/\epsilon^{\rm in}(q\neq 0,\w=0)$ does not vanish at non-vanishing wavevector. Therefore, an exciton could exist if its electron-hole distance is short enough. The screened Coulomb attraction $-W^{\rm in}$ in different approximations at $r_s=22$ \footnote{For the screened interaction in various approximations at $r_s \le 5$ see \cite{Kukkonen2021} and references therein.} is shown in Fig. \ref{fig:W}, as a function of wavevector (upper) and distance (bottom panel). In the RPA the interaction $-W^{\rm in}(r)$ decays rapidly with distance and is even slightly repulsive above $1/k_F$, but at short distances, below $r=1/k_F$, strong attraction sets in. As a consequence, GW-BSE could in principle yield  excitons with short e-h distance in metals, but to the best of our knowledge, this has not yet been explored in an \textit{ab initio} framework. We will therefore use BSE to investigate the short-range xc effects in the low-density HEG, and in particular, the possibility of formation of excitonic modes. 

\begin{figure}[ht]
\begin{center}
\includegraphics[angle=270,width=0.98\columnwidth]{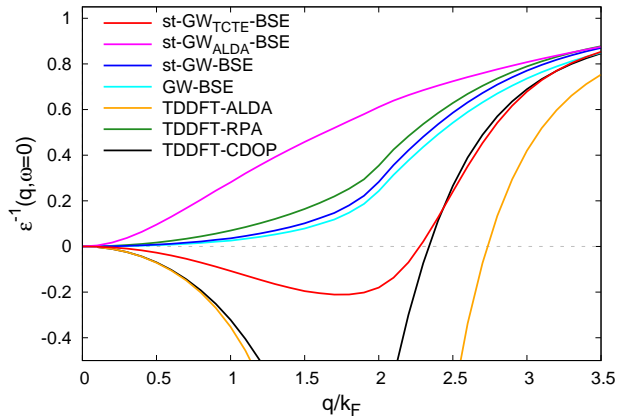}
\caption{Static inverse dielectric constant as a function of wavevector  
for $r_s=22$. The benchmark TDDFT-CDOP (black) is obtained in TDDFT with $f_{\rm xc}(q)$ in the interpolation of the QMC data of \cite{moroni1995a}  by Corradini {\it et al.} \cite{corradini1998a}. In green, TDDFT approximated in RPA. In orange, TDDFT in ALDA. In  cyan, BSE using a frequency-dependent RPA $W^{\rm in}$ to calculate the GW self-energy and static RPA $W^{\rm in}$ for the e-h interaction.  In blue, same but static RPA $W^{\rm in}$ in the GW self-energy. In magenta,  static ALDA $W^{\rm in}$ in the GW self-energy and in the e-h interaction. In red, BSE using the static TCTE $W^{\rm in}$ calculated with the ALDA kernel both in the GW self-energy and in the e-h interaction.}
\label{fig:static-screening}
\end{center}
\end{figure}

The signature of ghost excitations is the negative static screening, for which accurate Quantum Monte Carlo (QMC) predictions exist \cite{moroni1995a} \footnote{Also recent variational diagrammatic QMC results for high densities $r_s < 5$ \cite{Chen2019} agree with Ref. \cite{moroni1995a}. For recent QMC results in warm dense matter conditions see Refs. \cite{Dornheim2018,Dornheim2018b,Groth2019,Dornheim2022}.}. 
Fig. \ref{fig:static-screening} shows $1/\epsilon^{\rm out}(q,\omega=0)$ for $r_s =22$ [results for smaller $r_s$ can be found in the supplemental material  (SM) \cite{suppmat}]. The curve labeled TDDFT-CDOP was obtained in TDDFT [see line labeled 'Alternative' in scheme \eqref{eq:bse-scheme}]  using an $f_{\rm xc}(q,\omega=0)$  fitted to QMC results  \cite{moroni1995a} by Corradini {\it et al.} \cite{corradini1998a}; it should be an accurate benchmark.
The resulting $1/\epsilon^{\rm out}(q,\omega=0)$ has a minimum around $q=q_F$, where it is clearly negative. Approximating instead the TDDFT in the RPA, where $f_{\rm xc}$ is set to zero,  $1/\epsilon^{\rm out}$ is positive and 
overestimated for all $q$. It is now interesting to see what happens when the calculations are done using the BSE. Exact BSE should yield the exact TDDFT result, represented by CDOP. However, in spite of the fact that the RPA screened interaction is strongly attractive at short distances, our
GW-BSE with $\epsilon^{\rm in}=\epsilon^{\rm RPA}$  in (\ref{eq:bse-scheme})  does not show negative screening, which means that it completely misses the ghost physics.

\begin{figure}[ht]
\begin{center}
\includegraphics[angle=270,width=0.98\columnwidth]{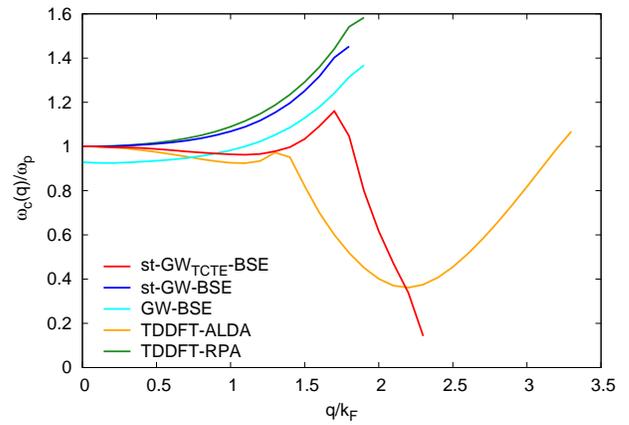}
\caption{Collective modes in the HEG at $r_s=22$ as a function of wavevector.
The lines indicate the position $\omega_c$ of  $\text{Re} \, \epsilon(q,\omega_c(q))=0$.  In green, TDDFT-RPA result. In orange, TDDFT-ALDA. 
In cyan, BSE using a frequency-dependent RPA $W^{\rm in}$ to calculate the GW self-energy and static RPA $W^{\rm in}$ for the e-h interaction, following scheme \eqref{eq:bse-scheme}.  In blue, same but static RPA $W^{\rm in}$ also for the GW self-energy. 
In red, BSE with static TCTE $W^{\rm in}$ calculated with the ALDA kernel, both in the GW self-energy and in the e-h interaction.  
}
\label{fig:collective-mode}
\end{center}
\end{figure}

The fact that GW-BSE has problems is confirmed by inspecting the collective mode at $\omega_c(q)$, which is defined by $\text{Re} \, \epsilon(q,\omega_c(q))=0$.
Fig. \ref{fig:collective-mode}  displays the dispersion $\omega_c(q)$ for $r_s =22$
(smaller $r_s$ are given in \cite{suppmat}). 
At low $q$ and high energy, the collective mode is the well known plasmon excitation \cite{Pines1952}. 
Starting from $q=0$, in the RPA the plasmon dispersion is always positive and quadratic \cite{Ichimaru1982,giuliani2005a}. In the GW-BSE this upwards dispersion is similar. 
However, an important exact constraint is the long wavelength limit of the plasmon energy, which should be  $\w_c(q\rightarrow0)=\w_p$ with $\w_p$ the classical plasma frequency, equal to the RPA result \cite{giuliani2005a,Mahan1981}.
Fig. \ref{fig:collective-mode} shows that GW-BSE violates this constraint \footnote{An analogous issue is the violation by standard BSE of the Goldstone condition $\omega_c(q\rightarrow0)=0$ for magnon excitations in ferromagnets  \cite{Mueller2016}.}. The origin of this problem is the inconsistent use of $\epsilon^{\rm in}$  in the standard GW-BSE as shown by scheme \eqref{eq:bse-scheme}, where a static approximation is made in the e-h interaction but not in the self-energy.  Instead, the correct $\w_c(q\rightarrow0)=\w_p$ is obtained in a st-GW-BSE approach, where the same static $\epsilon^{\rm in}$ is also used to calculate the self-energy \footnote{This is the static COulomb Hole plus Screened Exchange (COHSEX) approximation \cite{hedin1965a}}, as shown in Fig. \ref{fig:collective-mode}. This demonstrates the importance of a consistent use of $W$ in the BSE. 

Still, the st-GW-BSE calculation does not lead to    
 negative
$1/\epsilon^{\rm out}(q,\w=0)$ at any $q$, as shown in Fig. \ref{fig:static-screening}. 
This goes together with the absence of a low-energy mode in the st-GW-BSE result, which can be noticed in Fig. \ref{fig:collective-mode}. Such a low-energy mode, which could be the excitonic collective mode predicted by Takada \cite{takada2016a,Takada2005}, would make the negative density-density response function $\chi(q,\omega=0)$ stronger and could therefore lead to the desired negative  $\epsilon^{-1} = 1+v_c\chi$. We will  therefore search on the same footing for the origin of negative screening and of low-energy modes.

The clue to understand the underestimate of excitonic effects in GW-BSE is the effective screening of the e-h interaction. The use of the RPA for $W^{\rm in}$ is definitely a rough approximation, since 
the RPA is accurate  only for $q\to 0$ and for high densities \cite{giuliani2005a,Singwi1968}. One may therefore wonder whether the RPA might overscreen at short distances, and what happens when the e-h interaction is calculated with an improved screening. Since the accurate static screening is known to be negative over a large range of wavevectors, it is in particular exciting to investigate the impact of such a negative static screening on the e-h interaction and the excitons.  
To explore this, we calculate $W^{\rm in}$ using TDDFT beyond the RPA. By definition, using the CDOP $f_{\rm xc}(q,\omega=0)$ yields the correct $\chi(q,\omega=0)$. Interestingly, even the very simple and widely used  Adiabatic Local Density Approximation (ALDA), which corresponds to using the CDOP $f_{\rm xc}(q\to 0,\omega=0)$ at all $q$, reproduces the static screening and its negative part very closely for values of $q$ up to about $q\approx k_F$, while overestimating the effect of $f_{\rm xc}(q,\omega=0)$ at larger $q$, as can be seen in Fig. \ref{fig:static-screening}.
Indeed, the ALDA captures the peculiar features of the low-density HEG, with quantitative agreement at moderate densities and an overestimate of xc effects at very low densities \cite{panholzer2018a,Kaplan2022}. 
We will therefore use it for $W^{\rm in}$ in place of the RPA in order to show trends. The first interesting question is whether a negative $1/\epsilon^{\rm in}(q,\w=0)$ makes the effective Coulomb interaction attractive, which would lead to e-h repulsion. To answer,
Fig. \ref{fig:W} shows $-W^{\rm in}(r)$ in real space. In the RPA, $-W^{\rm in}(r)$ is oscillating with positive and negative regions, which can be associated with the Friedel oscillations of the screening charge density in real space \cite{Friedel1952,Harrison1980}. The inclusion of xc effects in ALDA merely enhances the amplitudes of these oscillations \footnote{Note that negative screening for a range of $q$ does therefore \textit{not} imply a globally attractive effective Coulomb interaction. This makes the interpretation of experimental observations, such as the structure factor of expanded liquid alkali metals \cite{Matsuda2007}, a non-trivial task.}. In other words, the negativity of $ 1/\epsilon^{\rm in} (q,\w=0)$ does not lead to qualitative changes in $-W^{\rm in}(r)$. 
Moreover, at short distances the absolute value of the screened interaction is reduced, rather than enhanced, with respect to the RPA. In other words, the exotic behavior of the screened interaction in reciprocal space does not lead to a globally attractive interaction
in real space, and the inclusion of xc effects simply leads to the weakening of the e-h interaction that is expected in ordinary semiconductors and insulators. As a consequence, when we now use $W^{\rm in}$ calculated in ALDA as input to st-GW-BSE, the resulting static $1/\epsilon^{\rm out}(q,\w=0)$, shown in Fig. \ref{fig:static-screening}, is even further from the benchmark CDOP than what was obtained using an RPA screened interaction in input. 

One might find this drastic failure of st-GW-BSE using an ALDA $W^{\rm in}$ surprising, since it would be quite intuitive to suppose that using a good approximation for the input screening would yield $W^{\rm in}(q,\omega=0)\approx W^{\rm out}(q,\omega=0)$. This intuition, however, supposes the GW approximation to yield a sufficient level of description, while it suffers from known problems, in particular self-polarization \cite{Nelson2007,Romaniello2009,Aryasetiawan2012,Chang2012}. Indeed, independently of the approximation that is used, the screening in $W$ expresses the classical electrostatic potential induced by the charge response, while electrons and holes should also experience an induced xc potential. Such an induced xc potential eliminates the self-polarization and weakens the effective screening especially at short range. It can be taken into account approximately by using a Test-Charge-Test-Electron (TCTE) inverse dielectric function \cite{streitenberger1984a,Mahan1989,delsole1994a} $\epsilon_{\rm TCTE}^{-1}=1+(v_c+f_{\rm xc})\chi$  instead of 
$\epsilon^{-1}=1+v_c\chi$ to screen $W^{\rm in}$. 
Contributions from vertex corrections such as ladder diagrams  
become more important for lower densities  \cite{fetter2003a,Yasuhara1974,Mattuck1992,Irmler2019}. Using a TCTE screening for $W^{\rm in}$ corresponds to the inclusion of a vertex correction that is derived from
a local approximation of the self-energy \cite{delsole1994a,Bruneval2005}. Such an approximation for the vertex correction is justified by the fact that 
 the self-energy is expected to be of short range in the HEG.
 
We therefore recalculate $W^{\rm in}(q,\omega=0)$ with a TCTE screening based on the ALDA $f_{xc}$. Fig. \ref{fig:static-screening} shows that $-W_{\rm TCTE-ALDA}$ is very different from $-W_{\rm ALDA}$. In reciprocal space, it develops a pronounced dip at $q \sim 2 k_F$, in correspondence to a very weak feature in $W_{\rm RPA}$. This dip is consistent with the shoulder found in \cite{richardson1994a} for higher density \footnote{Interestingly, $f_{\rm xc}(q\to 0,\omega=0)$ seems to be an approximate average of the TCTE $-W^{\rm in}$, which explains why both quantities used as e-h interaction produce a similar effect.}.
In real space, at short distances $r$ the TCTE screened interaction $W_{\rm TCTE-ALDA}(r)$ is closer than $W_{\rm ALDA}(r)$ to the bare Coulomb interaction $v_c(r)$, and even stronger than $v_c(r)$ below about $1/(2k_F)$.
Using now $W^{\rm in}=W_{\rm TCTE-ALDA}$, this enhanced short-range interaction leads to a drastic improvement of the st-GW$_{\rm TCTE}$-BSE result. Most importantly, we find negative screening, as shown in Fig. \ref{fig:static-screening}, although the effect is underestimated with respect to the benchmark CDOP result. This is also apparent at smaller $r_s$, shown in \cite{suppmat}. Still, the description is qualitatively convincing:  Fig. \ref{fig:collective-mode} shows that while the $q\to 0$ plasmon limit remains exact, st-GW$_{\rm TCTE}$-BSE now leads to a negative plasmon dispersion, which is an expected exchange-correlation effect \cite{Ichimaru1982,giuliani2005a}. Moreover, an
abrupt decrease of the energy of the collective mode slightly before $q=2k_F$ appears: the mode drops by more than a factor of 5 up to about $q=2.4 k_F$, before it is damped out. 

\begin{figure*}[ht]
\begin{center}
\includegraphics[angle=270,width=0.96\columnwidth]{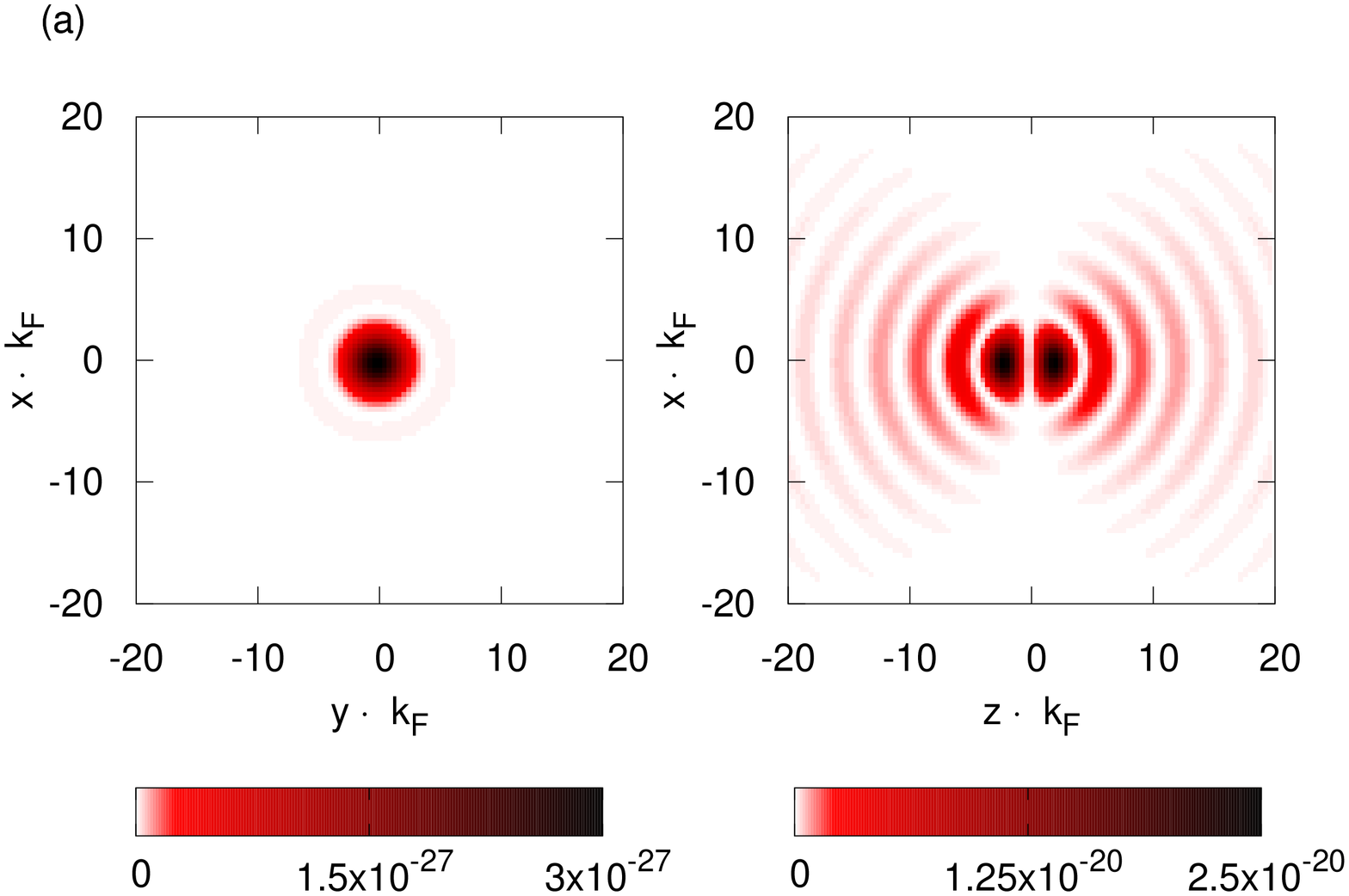}
\includegraphics[angle=270,width=0.96\columnwidth]{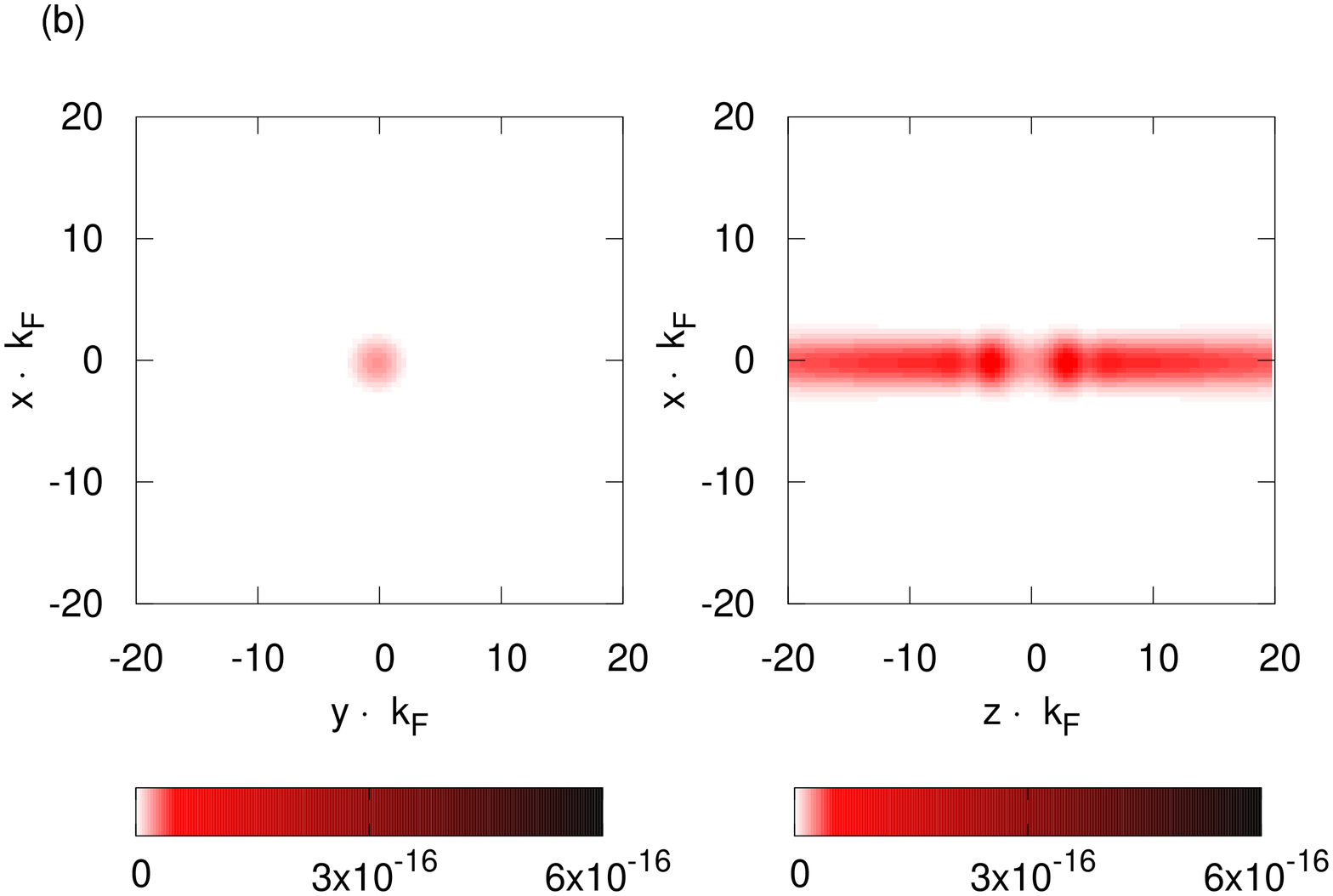}
\includegraphics[angle=270,width=0.96\columnwidth]{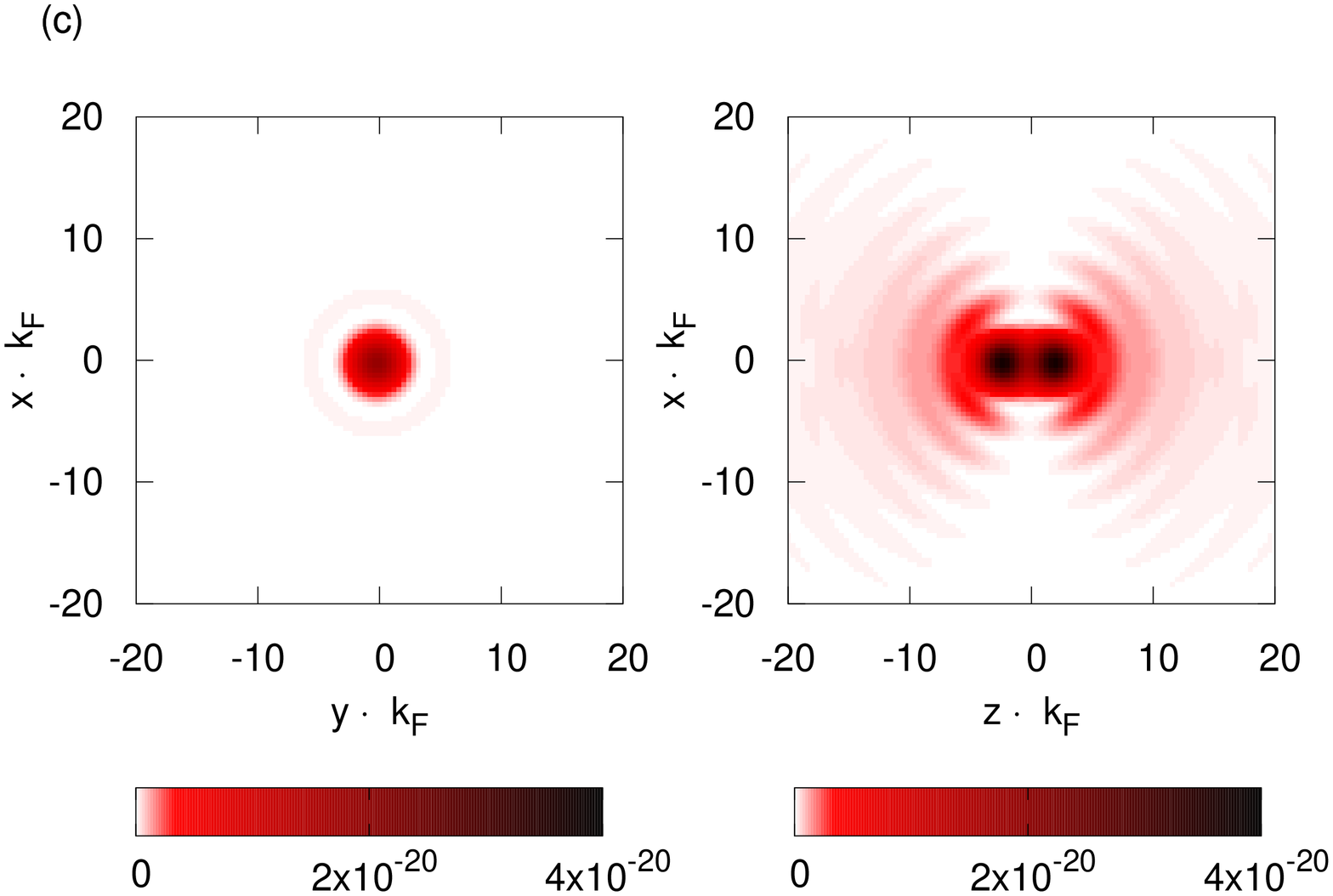}
\includegraphics[angle=270,width=0.96\columnwidth]{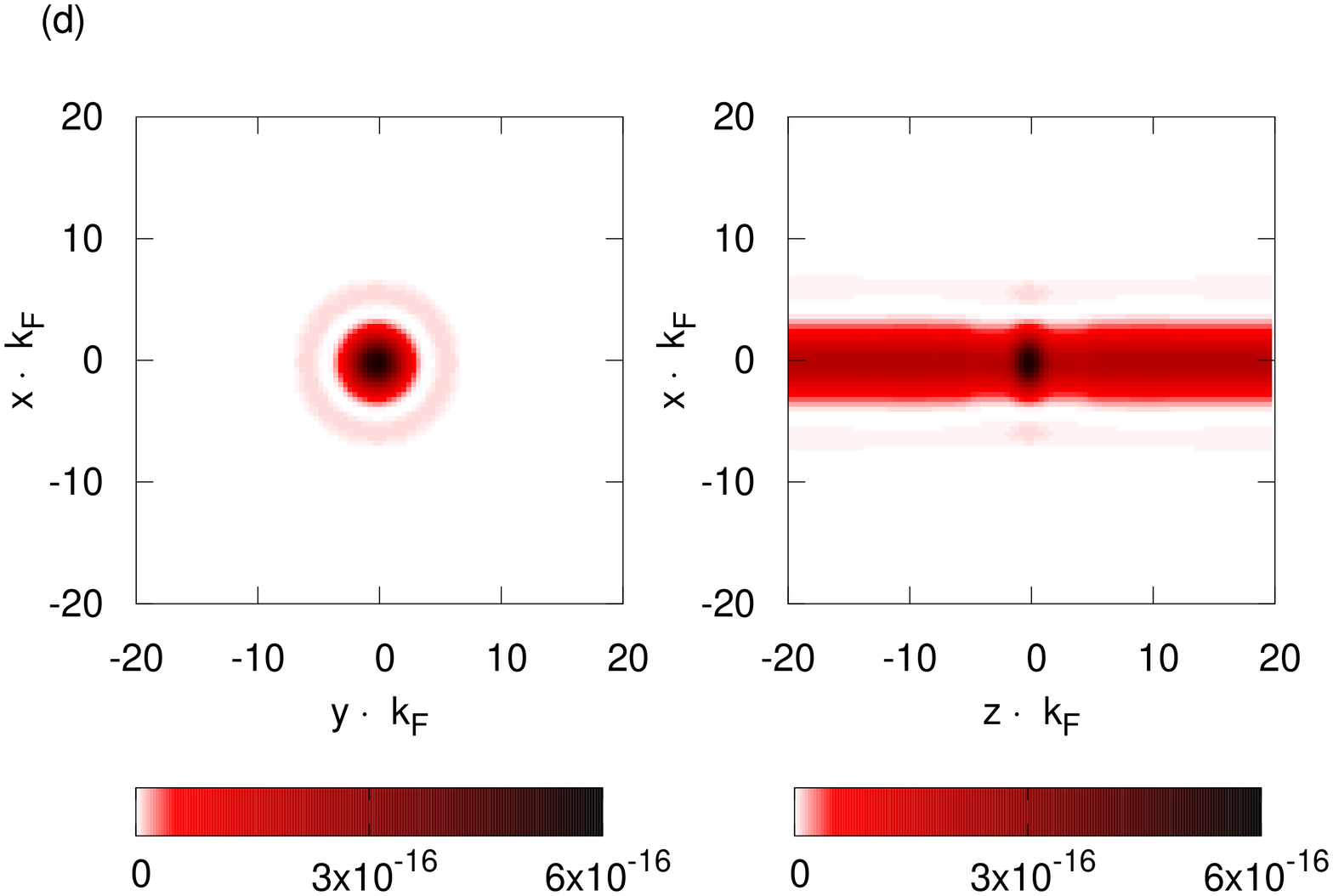}
\caption{Absolute square of the correlated e-h amplitude obtained from the BSE or TDDFT Casida-equation for the HEG at $r_s=22$. We choose ${\bf q}=q\hat{\bf z}$. In each panel: Left side, xy plane; right side, yz plane. 
(a) Plasmon at $q=0.001k_F$ calculated in the ALDA. 
(b)  Excitonic collective mode at $q=2k_F$ from the BSE using the static TCTE $W^{\rm in}$ calculated with the ALDA kernel.
(c) Ghost exciton modes obtained from the ALDA for the irreducible polarizability at $q=0.001k_F$: sum of the two imaginary poles at energies $\pm  7.90i \times 10^{-6}$ Hartree. (d) Excitonic collective mode at $q=2k_F$ from the TDDFT Casida-equation in ALDA. 
}
\label{fig:correlation}
\end{center}
\end{figure*}

With a physically reasonable approximation to the BSE at hand that describes the important qualitative features, and in particular the appearance of a low-energy mode,  we can now discuss the character of that mode. First, the origin of the phenomenon is clearly the e-h attraction: no such mode appears when this interaction is switched off in the calculation.  
The mode is hence of excitonic nature. To dig deeper, Fig. \ref{fig:correlation} shows for $r_s=22$ the e-h amplitudes $|\Psi_{\lambda^q}(\bfr)|^2$, where $\Psi_{\lambda^q}(\bfr)$ are the eigenfunctions of the e-h BSE hamiltonian which in the HEG can depend only on the e-h distance $r$ and its direction relative to the momentum ${\bf q}$ of the excitation. For independent electrons and holes this amplitude would be a constant, whereas excitonic effects usually appear as a localization of the electron cloud around the hole \cite{Rohlfing2000,martin2016a}. For Fig. \ref{fig:correlation}, the excitation $\lambda^q$ is chosen to be the collective mode. At low $q$ this is the plasmon, for which the e-h amplitudes are shown in Fig. \ref{fig:correlation}(a), in a plane perpendicular to ${\bf q}$ (left panel) and with ${\bf q}$ in plane in $z$ direction (right panel), respectively. The plasmon amplitude is different from the RPA result of
Egri \cite{Egri1983,Egri1985}, mainly because only the resonant part of the amplitude is shown in that work. Here, instead, we take all resonant and anti-resonant contributions into account, which leads to strong cancellations and an overall oscillating, but essentially vanishing, amplitude at this small q. 
Fig. \ref{fig:correlation}(b) shows the low-energy collective mode found for  $q=2k_F$ at $E_{\lambda}=6.627 \times 10^{-3}$ Hartree in st-GW$_{\rm TCTE}$-BSE with $W^{\rm in}_{\rm TCTE-ALDA}$. As the scale shows, {the amplitude of the collective mode found at this larger $q$  is  significant.  In a plane perpendicular to ${\bf q}$ its} distribution is spherical and localized around the origin. However, it is much more delocalized in direction of ${\bf q}$,  which makes it very anisotropic: indeed, it looks like an unbound state along ${\bf q}$, but in the two perpendicular directions it is localized like a strongly bound exciton. 
We also show in Fig. \ref{fig:correlation}(c) the ghost excitation mode, obtained as poles of $\epsilon^{\rm out}$, for the same small $q$ as the plasmon. Interestingly, its shape  is quite similar to that of the plasmon. 
Note that here we show the sum of the two imaginary twin ghost poles at energies $\pm i|E|$.
Each mode alone breaks the symmetry $+\bfq \leftrightarrow -\bfq$ (i.e., right $\leftrightarrow$ left in the picture), but the symmetry is restored by their sum. 

This wavefunction analysis shows that the character of the ghost excitation is rather that of a plasmon, as proposed by \cite{dolgov1981a,takayanagi1997a}, than that of an exciton as 
Takada's terminology \cite{Takada2005,takada2016a} might suggest. Instead, the observable low-energy mode in the density-density response function shows indeed excitonic characteristics, at least in two of the three dimensions. This is not in contradiction with the fact that both modes are exclusively due to strong electron-hole attraction: it merely shows that we have to define carefully what we mean by ``an exciton''.

A final intriguing question 
 is the interpretation of the same modes obtained directly  from a TDDFT calculation [indicated as alternative in scheme \eqref{eq:bse-scheme}]: 
  Fig. \ref{fig:collective-mode} shows that at wavevectors around $q=2k_F$ also the ALDA yields a low-energy mode, which is associated with  the appearance of the negative static screening in TDDFT-ALDA observed in Fig. \ref{fig:static-screening}. We therefore 
 solve the Casida equation \cite{Casida1995}, which formulates TDDFT in a basis of independent-particle transitions and therefore yields e-h amplitudes.
The results $|\Psi^{\rm ALDA}_{\lambda^q}(\bfr)|^2$ for ALDA are shown in Fig. \ref{fig:correlation}(d). The wavefunction is strikingly similar to the one resulting from the BSE, although TDDFT and BSE are two completely different approaches. Indeed, while exact TDDFT and exact BSE must yield the same density-density response function, we are not aware of any proof that they should also yield the same e-h wavefunction. The main difference found in our approximate calculations is that in TDDFT the intensity is higher. This is consistent with the fact that st-GW$_{\rm TCTE}$-BSE with the TCTE e-h interaction underestimates the negative screening, while TDDFT-ALDA has a tendency to overestimate, as shown by Figs. \ref{fig:static-screening} and \ref{fig:collective-mode}. Still, the agreement is surprisingly good, also in view of the simple approximations made here, and it gives strong confidence in our observations and qualitative understanding.

In conclusion, while simple models \cite{knox1963a,bassani1975a} 
predict that excitons do not exist in metals due to perfect macroscopic screening,the existence of bound electron-hole pairs and intriguing phenomena such as imaginary poles in the dielectric function called ghosts, or negative static screening, are made possible due to the imperfect screening of the electron-hole interaction at short distances.  Such an imperfect screening at short distances \cite{Mahan1981,fetter2003a,giuliani2005a} occurs already for the interaction between classical charges, but the effect is not strong enough to yield excitons from the Bethe-Salpeter equation. Instead, the e-h interaction must be screened by a test charge-test-electron dielectric function, which takes into account the fermionic character of the charges, and which excludes self-polarization. Using such an effective interaction, even with simple approximations, in the BSE captures the qualitative picture correctly. It yields ghosts and low-energy excitations leading to negative static screening at low densities, which allows us to assign a plasmonic character to the ghost excitation, while we find two-dimensional exciton binding features for the low-energy mode. Still, our st-GW$_{\rm TCTE}$-BSE approximation underestimates the strength of the negative screening. The correction might be further increased by dynamical effects that are neglected in the current approximations for the e-h interaction \cite{Strinati1982}, which would reflect the fact that it takes time to build up a screening cloud \cite{canright1988a,alducin2004a,schoene2000a}, allowing a bound exciton with short e-h distance to stabilize further. On the other hand, the strong dispersion of the low-energy mode 
  suggests significant cancellations of dynamical effects \cite{Cudazzo2020}, which justifies the static approximation. To get a more precise estimate is a complex task
\cite{Rohlfing2000,Ma2009,Spataru2013,Gao2016,Loos2020} which is beyond the scope of the present work. 

Interestingly, the ghost and collective exciton physics is also found without solving the BSE, but using TDDFT in the simple ALDA.  This may seem puzzling, since ALDA is known to completely miss excitonic effects in semiconductors and insulators \cite{Botti2007}. However, our analysis shows that this is not a coincidence for wrong reasons, since the e-h wavefunctions obtained from ALDA and from the BSE are similar. The success of ALDA may be explained by the fact that in the present case the important physics happens at short distances, but it cannot be simply transposed to real materials. Instead, the solution of the BSE using a TC-TE effective e-h interaction should be valid also for the cases that require a better treatment of the long range. This opens the way for a broad field of potential applications, searching for ghost excitons and the consequent intriguing many-body effects in systems such as doped or photoexcited semiconductors \cite{Mott1968,Shah1977,Schleife2011,Steinhoff2017,Siday2022}, or low-density electron gases at surfaces and interfaces \cite{Ohtomo2004,Santander-Syro2011,Lin2013}.

\begin{acknowledgments}
We thank Martin Panholzer for fruitful discussions.  This work is supported by a public grant overseen by the French National Research Agency (ANR) as part of the ``Investissements d’Avenir'' program (Labex NanoSaclay, reference: ANR-10-LABX-0035), by the Magnus Ehrnrooth Foundation, and by the European Research Council (ERC Grant Agreement n. 320971).
  The research leading to these results has received funding also from the People Programme (Marie Curie Actions) of the European Union’s Seventh Framework Programme (FP7/2007-2013) under REA grant agreement n. PCOFUND-GA-2013-609102, through the PRESTIGE programme coordinated by Campus France. Computational
time was granted by GENCI (Project No. 544).   
\end{acknowledgments}

\bibliography{heg}

\end{document}


\title{Supplemental material for \\
Short-range excitonic phenomena in low-density metals} 

\newcommand{\lsi}{LSI, CNRS, CEA/DRF/IRAMIS, \'Ecole Polytechnique, Institut Polytechnique de Paris, F-91120 Palaiseau, France}
\newcommand{\etsf}{European Theoretical Spectroscopy Facility (ETSF)}
\newcommand{\soleil}{Synchrotron SOLEIL, L'Orme des Merisiers, Saint-Aubin, BP 48, F-91192 Gif-sur-Yvette, France}

\author{Jaakko Koskelo}
\affiliation{\lsi}
\affiliation{\etsf}

\author{Lucia Reining}
\affiliation{\lsi}
\affiliation{\etsf}

\author{Matteo Gatti}
\affiliation{\lsi}
\affiliation{\etsf}
\affiliation{\soleil}

\maketitle

\section{Computational details}

We have implemented the BSE in a computer code that explicitly exploits: i) the fact that QP wavefunctions in the HEG are plane waves and ii) the cylindrical symmetry of the problem: integrating over the angular variable in the BSE hamiltonian decreases the 3 momentum components to 2.
In this way, we can afford the dense $\bfk$ point sampling required for the HEG.
GW calculations have been done using adaptive integration routines for the momentum and frequency integrals.
For BSE calculations, the momentum grid spacing varies as a function of momentum transfer so that hamiltonian is of roughly the same size for all the momentum transfers. For example, the momentum spacing along the axes is $k_F/2061$ and $k_F/57$ for $q=0.001k_F$ and $q \geq 2.0k_F$, respectively.

\section{Results for $r_s=4$ and $8$}

In order to observe trends, we present in the following results for the main points discussed in the article, but for higher densities $r_s=4$ and $8$. The qualitative trends for $r_s=4$ and $r_s=8$ are similar as those observed for $r_s=22$ in main text, a part from the fact that for $r_s=4$ the static screening is always positive. The quantitative differences between the various approximations increase with increasing $r_s$, as an effect of increasing impact of the Coulomb interaction.

\begin{figure}[ht]
\begin{center}
\includegraphics[angle=270,width=0.49\columnwidth]{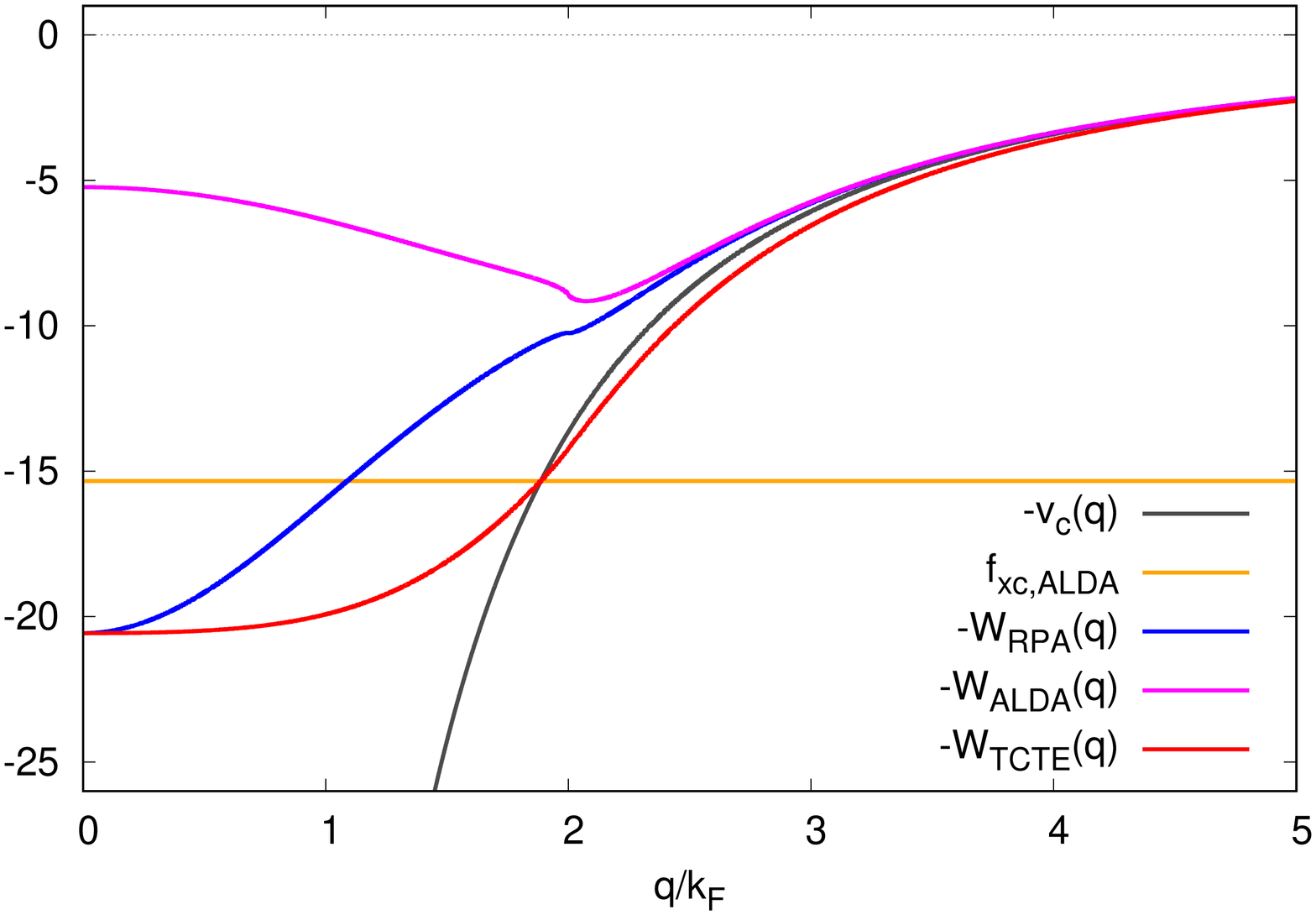}
\includegraphics[angle=270,width=0.49\columnwidth]{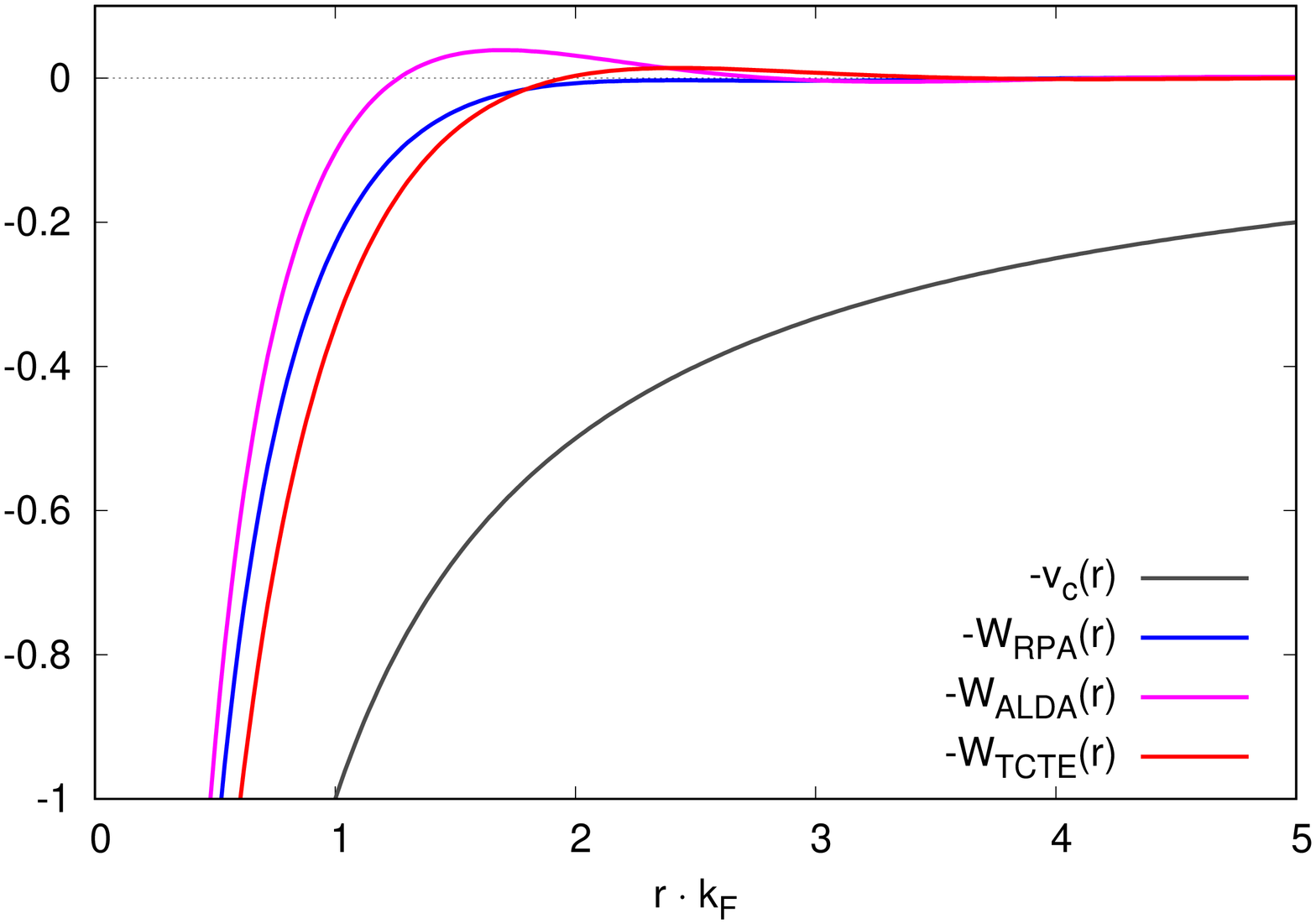}
\caption{Same as Fig. 1 in the main text, for $r_s=4$. 
}
\label{fig:W-supp-4}
\end{center}
\end{figure}

\begin{figure}[ht]
\begin{center}
\includegraphics[angle=270,width=0.49\columnwidth]{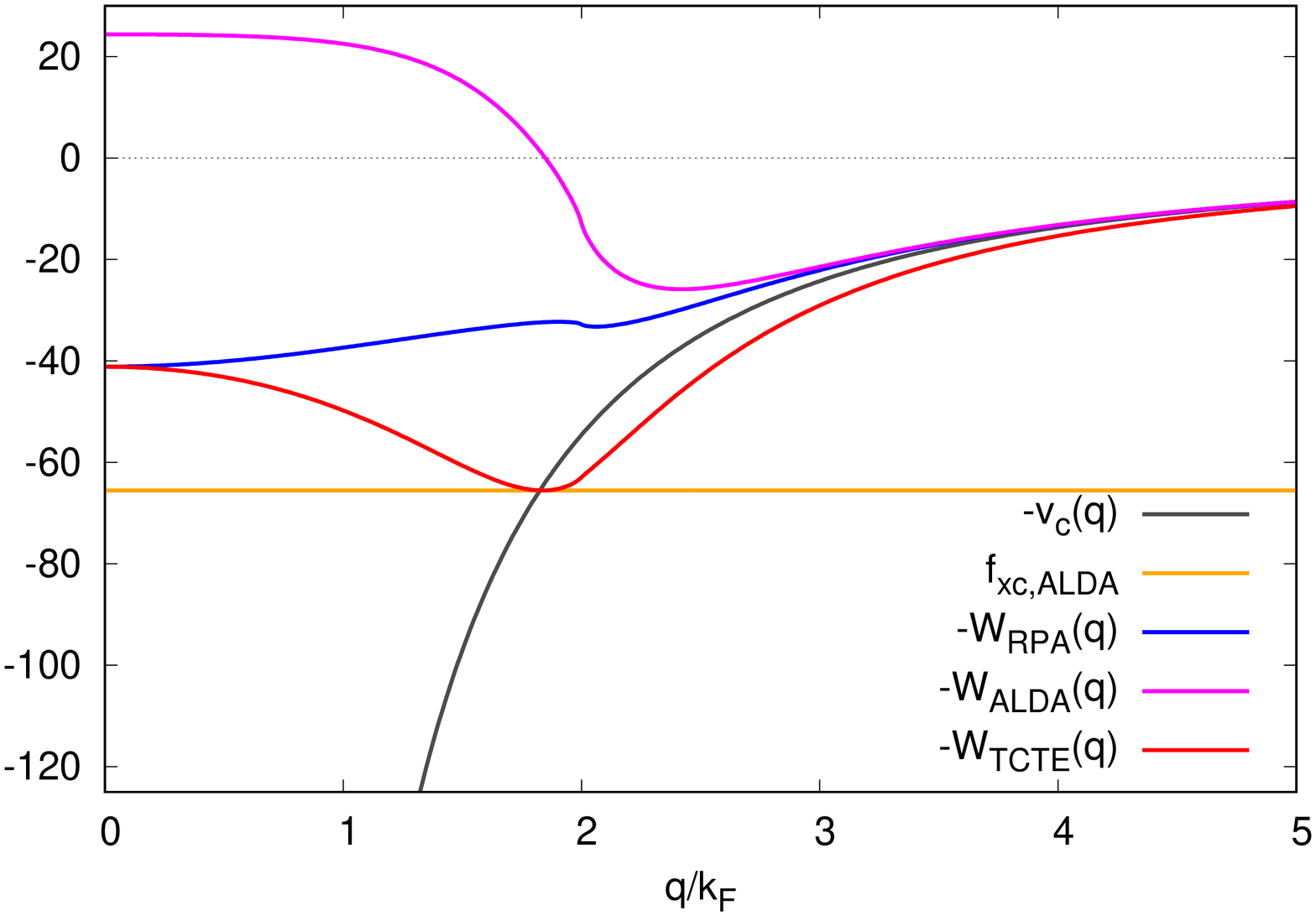}
\includegraphics[angle=270,width=0.49\columnwidth]{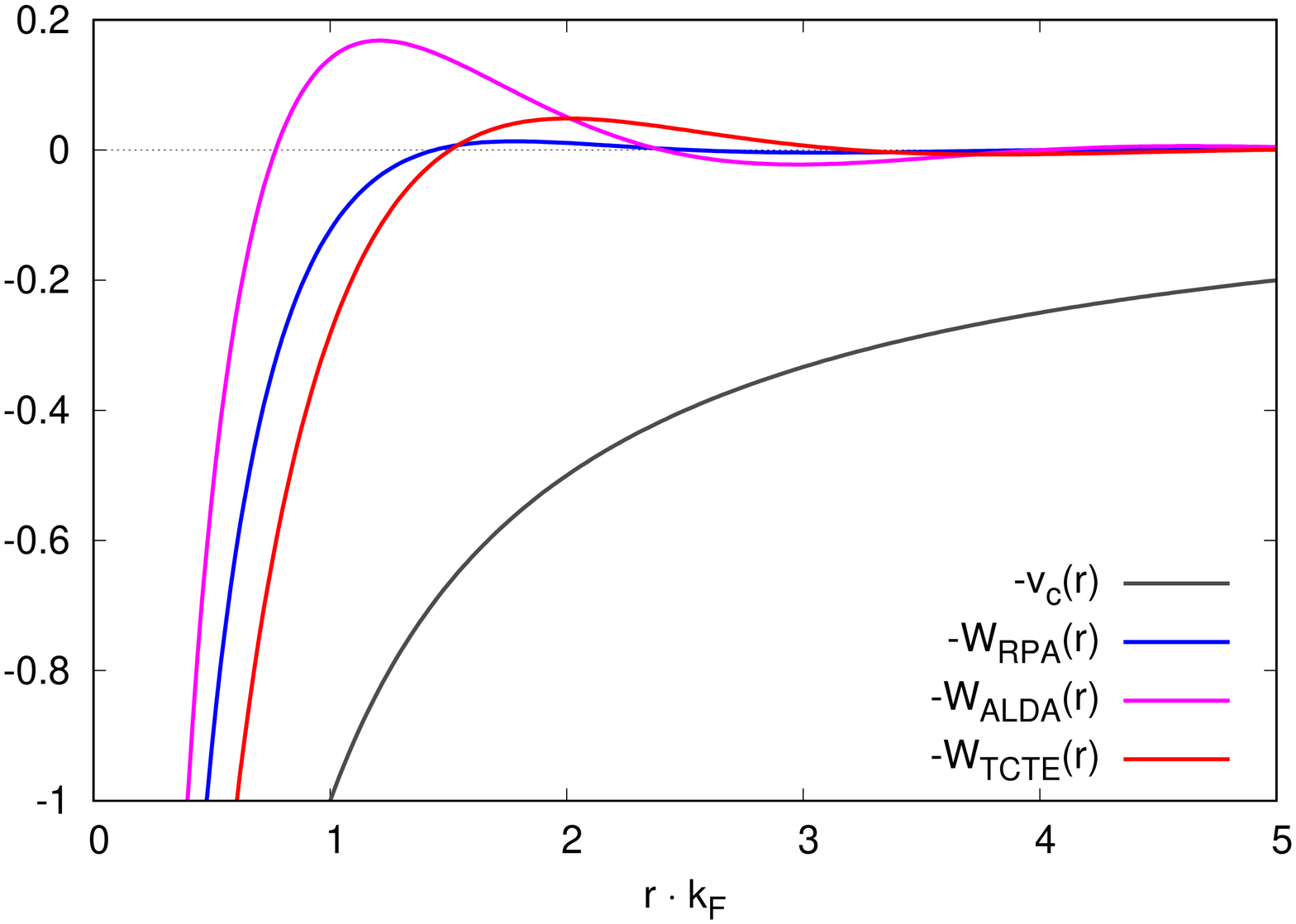}
\caption{Same as Fig. 1 in the main text, for $r_s=8$. 
}
\label{fig:W-supp-8}
\end{center}
\end{figure}

The screening of the Coulomb interaction, see Figs. \ref{fig:W-supp-4}-\ref{fig:W-supp-8}, is not complete at short distances $r < 1/k_F$, where $-W_{\rm TCTE}(r) < -W_{\rm RPA}(r) < -W_{\rm ALDA}(r)$. Contrary to $r_s=22$, $-W_{\rm TCTE}(r)>-v_c(r)$ for all $r$.
In all approximations and densities, the screened interactions are oscillatory functions of $r$. In reciprocal space, at $r_s=8$, while  $-W_{\rm ALDA}(q)$ becomes positive (i.e., repulsive) for small $q$, -$W_{\rm TCTE}(q)$ is always negative, with a pronounced minimum at $q \sim 2 k_F$, similarly to $r_s=22$ in the main text, although the effect is weaker.

\begin{figure}[ht]
\begin{center}
\includegraphics[angle=270,width=0.49\columnwidth]{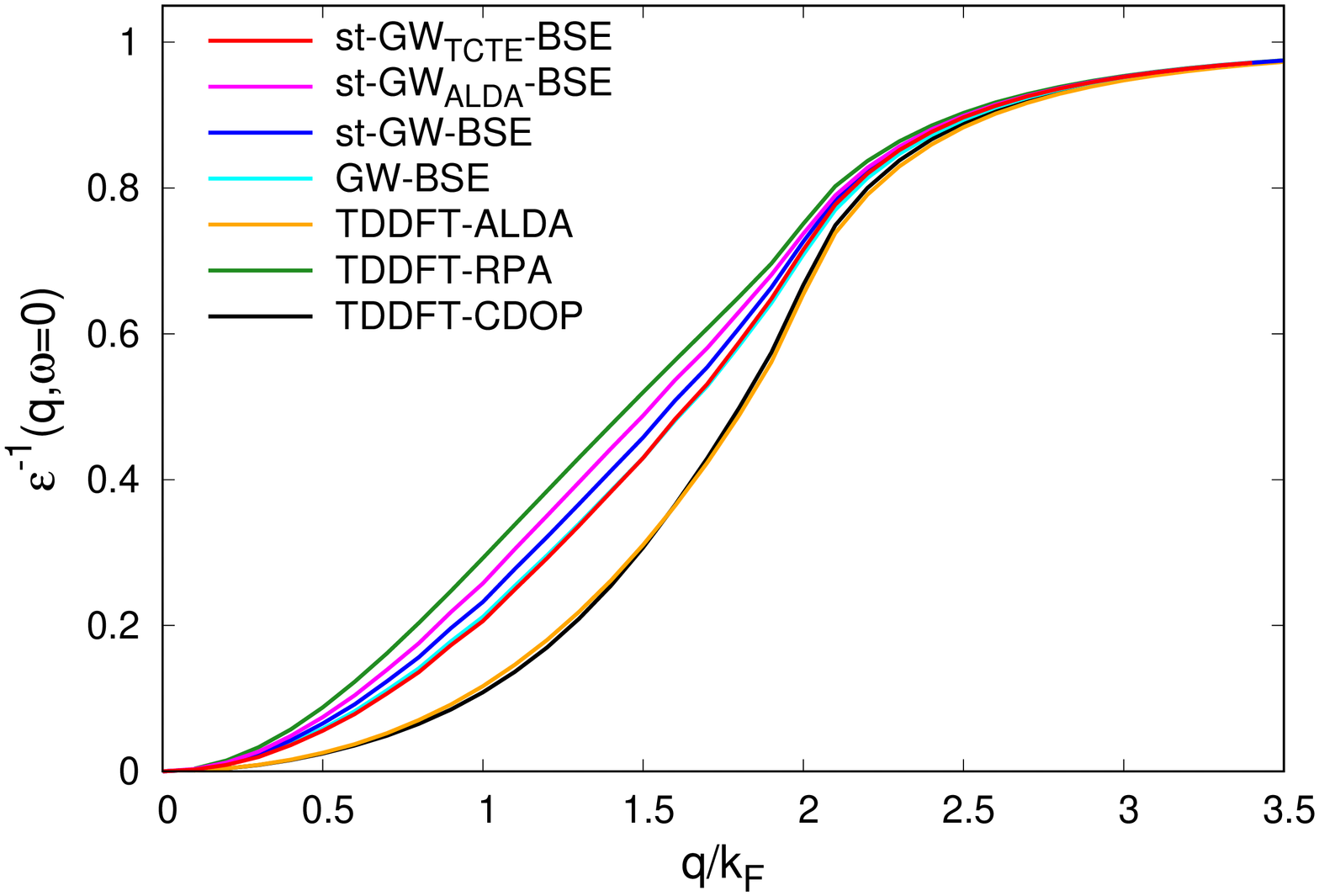}
\includegraphics[angle=270,width=0.49\columnwidth]{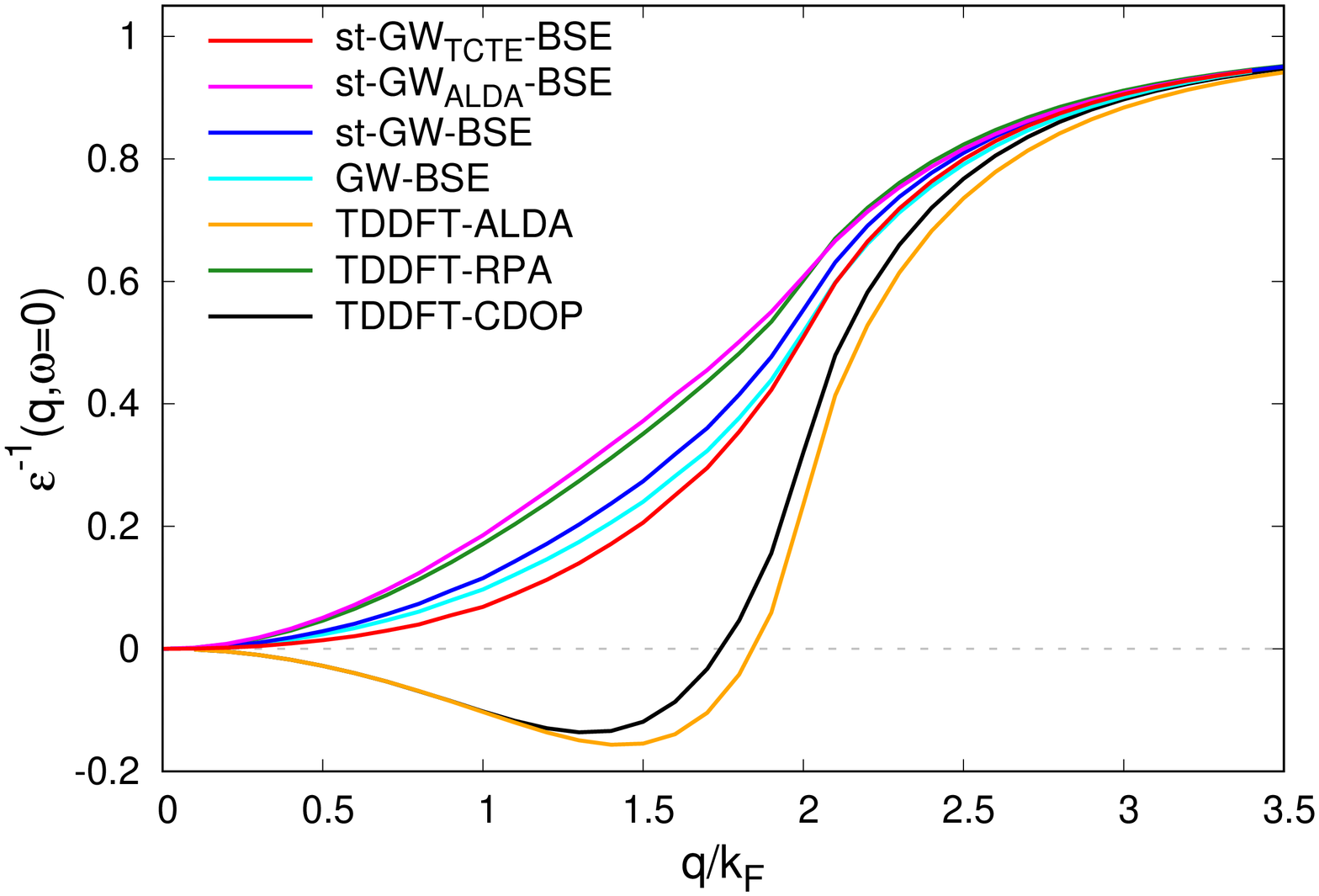}
\caption{Same as Fig. 2 in the main text, for $r_s=4$ (left) and $r_s=8$ (right). }
\label{fig:static-screening-supp-4-8}
\end{center}
\end{figure}

Concerning the static screening $1/\epsilon^{\rm out}(q,\w=0$), see Fig. \ref{fig:static-screening-supp-4-8}, 
TDDFT-ALDA results are similar to the benchmark TDDFT-CDOP, although some discrepancies are visible at $r_s=8$ for $q>k_F$. For both $r_s=4$ and $r_s=8$, the BSE using the static TCTE $W^{\rm in}$ calculated with the ALDA kernel both in the GW self-energy and in the e-h interaction gets closer to the benchmark than the BSE that uses a static RPA $W^{\rm in}$ everywhere. However, contrary to the benchmark  TDDFT-CDOP it does not yield a negative screening at $r_s=8$.

\begin{figure}[ht]
\begin{center}
\includegraphics[angle=270,width=0.49\columnwidth]{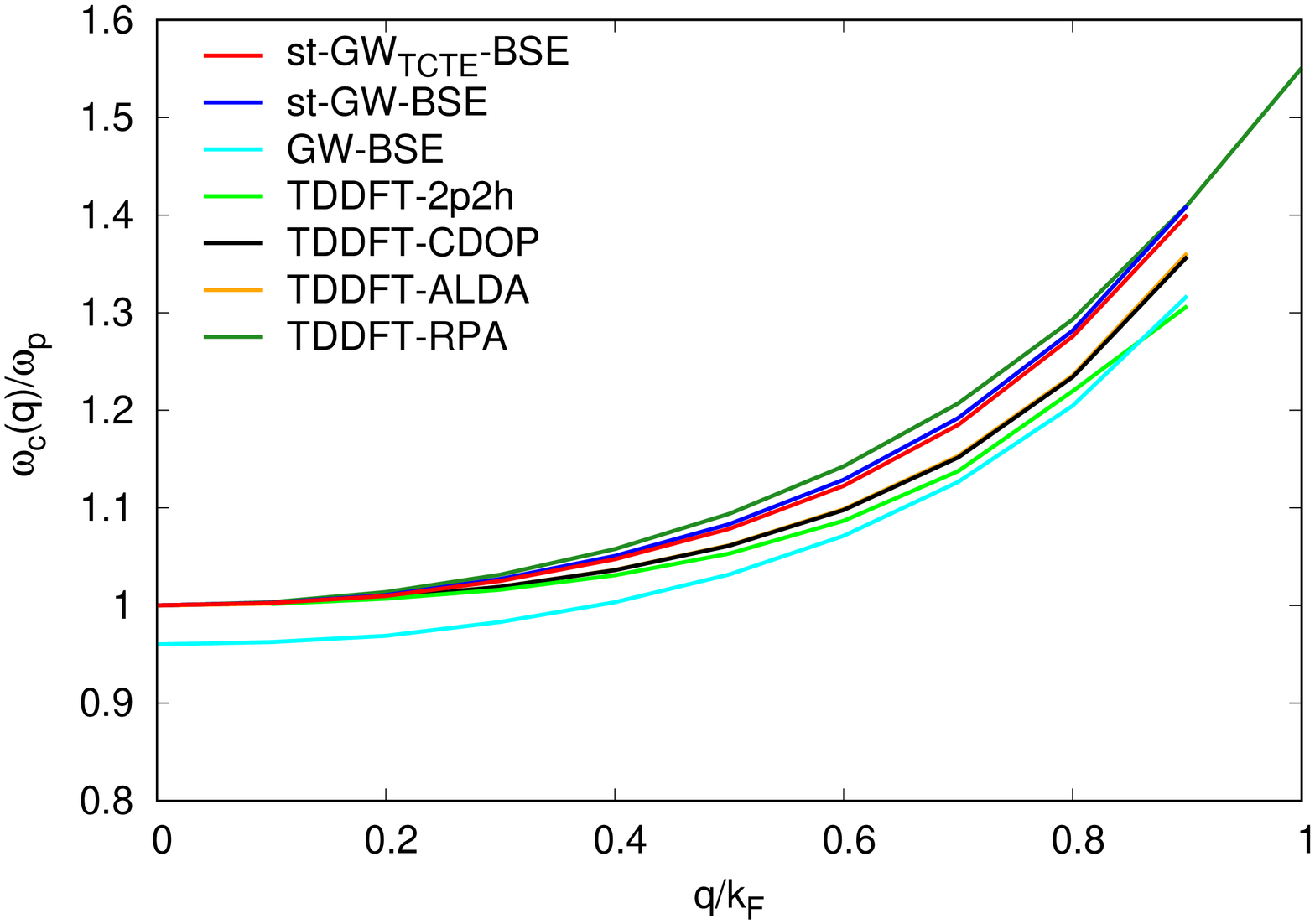}
\includegraphics[angle=270,width=0.49\columnwidth]{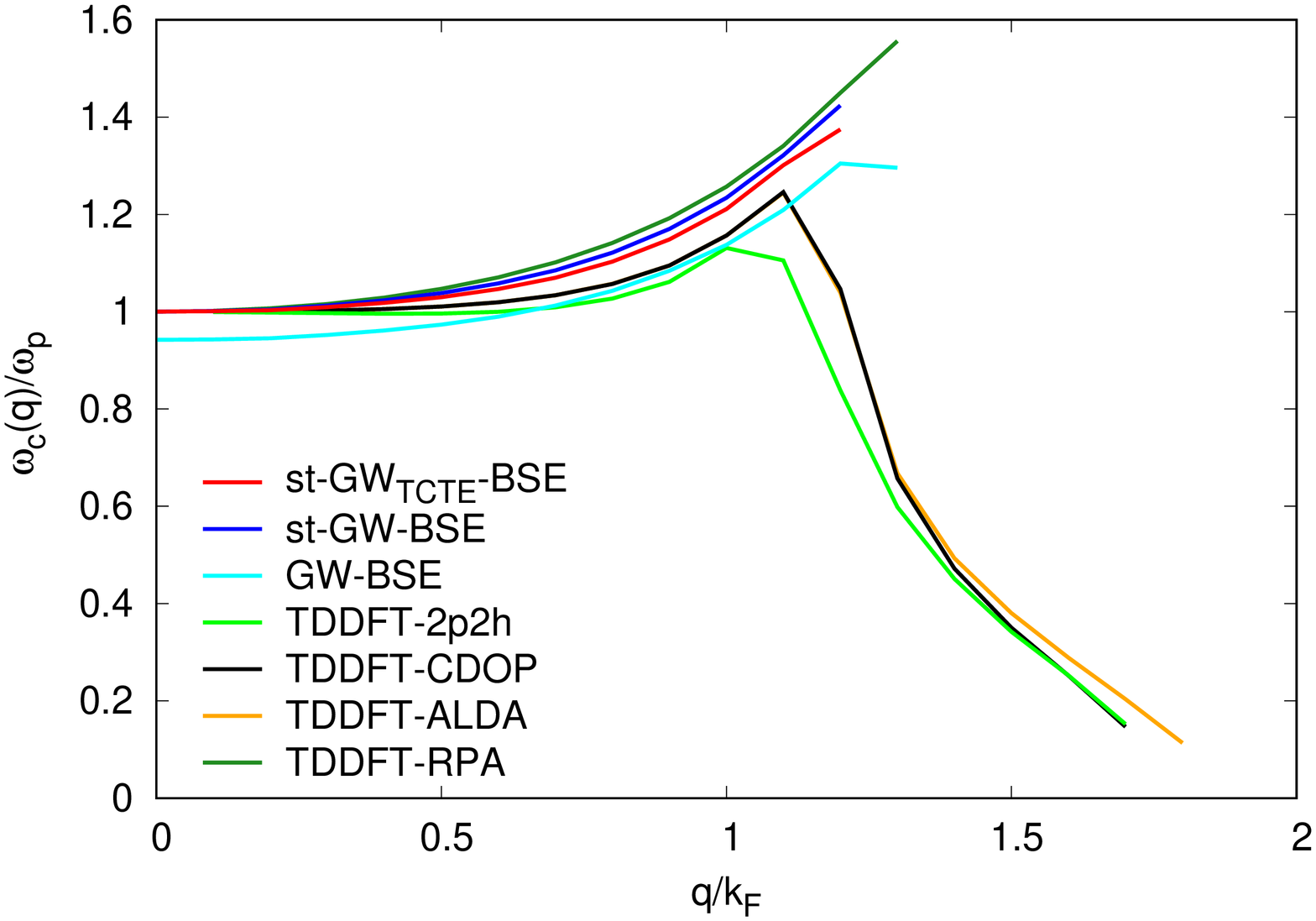}
\caption{Same as Fig. 3 in the main text, for $r_s=4$ (left) and $r_s=8$ (right). 
Moreover the results of TDDFT calculations using the CDOP and 2p2h kernels are given in black and light green, respectively.}
\label{fig:collective-mode-supp-4-8}
\end{center}
\end{figure}

The dispersion $\w_c(q)$ of the collective modes as a function of wavevector $q$, corresponding to the condition $\text{Re}\, \epsilon(q,\w_c(q))=0$, is shown in Fig. \ref{fig:collective-mode-supp-4-8}.QMC benchmark results are available only for static (i.e., $\w=0$) properties of the HEG. Therefore, with respect to Fig. 3 in the main text, besides the (static and local) ALDA, here we have also added TDDFT results using  the dynamical and nonlocal two-particle two-hole (2p2h) kernel from Ref. \cite{panholzer2018a} (obtained from the accurate correlated equations of motion approach of B\"ohm {\it et al.} \cite{bohm2010a}). We consider the 2p2h results (available for $r_s \le 8$)
as the benchmark for the dynamical properties of the HEG. We have also added TDDFT results using the static and nonlocal CDOP kernel: while these results are the benchmark at $\omega=0$, where they coincide with the 2p2h results, they have no reason to be accurate at non-vanishing frequencies.

At low $q$ and large energies, the collective modes are the plasmons.
While at $r_s=4$ the plasmon has a quadratic upwards dispersion before it decays into the e-h continuum in all approximations, at $r_s=8$ it flattens in all the TDDFT approximations beyond the RPA. The standard GW-BSE violates the condition $\w_c(q\to0)=\w_p^0$ (see main text) at all densities.
At $r_s=8$ TDDFT using the 2p2h kernel predicts the excitonic collective mode  with negative dispersion  at large $q> k_F$. TDDFT using the ALDA or CDOP kernels yields a very similar result, with CDOP and ALDA being slightly different only at large $q$, as one would expect.   Instead, the low-energy collective mode is missed by the BSE in all the approximations, consistently with the lack of negative screening in the BSE result at this density, and with the underestimation of the effect by the BSE found at $r_s=22$ in the main text.

\section{Results at $r_s=22$ using the CDOP $f_{\rm xc}$}

\begin{figure}[ht]
\begin{center}
\includegraphics[angle=270,width=0.49\columnwidth]{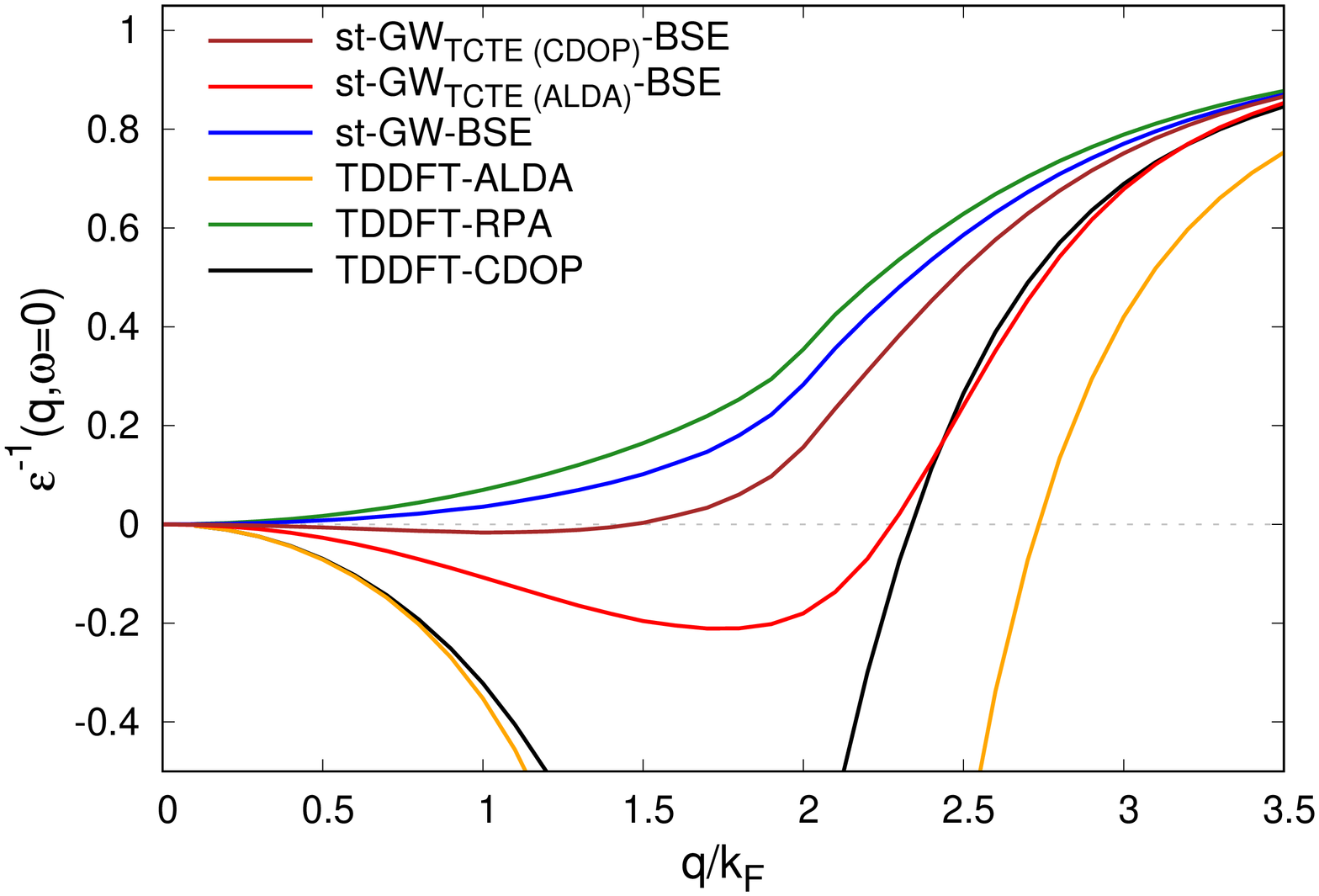}
\caption{Static inverse dielectric constant as a function of wavevector 
for $r_s=22$. The benchmark TDDFT-CDOP (black) is obtained in TDDFT with $f_{\rm xc}(q,\w=0)$ in the interpolation of the QMC data of \cite{moroni1995a}  by Corradini {\it et al.} \cite{corradini1998a}. In green, TDDFT approximated in RPA, where $f_{\rm xc}=0$. In orange, TDDFT with $f_{\rm xc}$ in ALDA. In blue, BSE using a static RPA $W^{\rm in}$ in the GW self-energy and for the e-h interaction.  In red, BSE using the static TC-TE $W^{\rm in}$ calculated with the ALDA kernel both in the GW self-energy and in the e-h interaction.
In brown, BSE using the static TCTE $W^{\rm in}$ calculated with the CODP kernel both in the GW self-energy and in the e-h interaction.}
\label{fig:static-screening-cdop}
\end{center}
\end{figure}

\begin{figure}[ht]
\begin{center}
\includegraphics[angle=270,width=0.49\columnwidth]{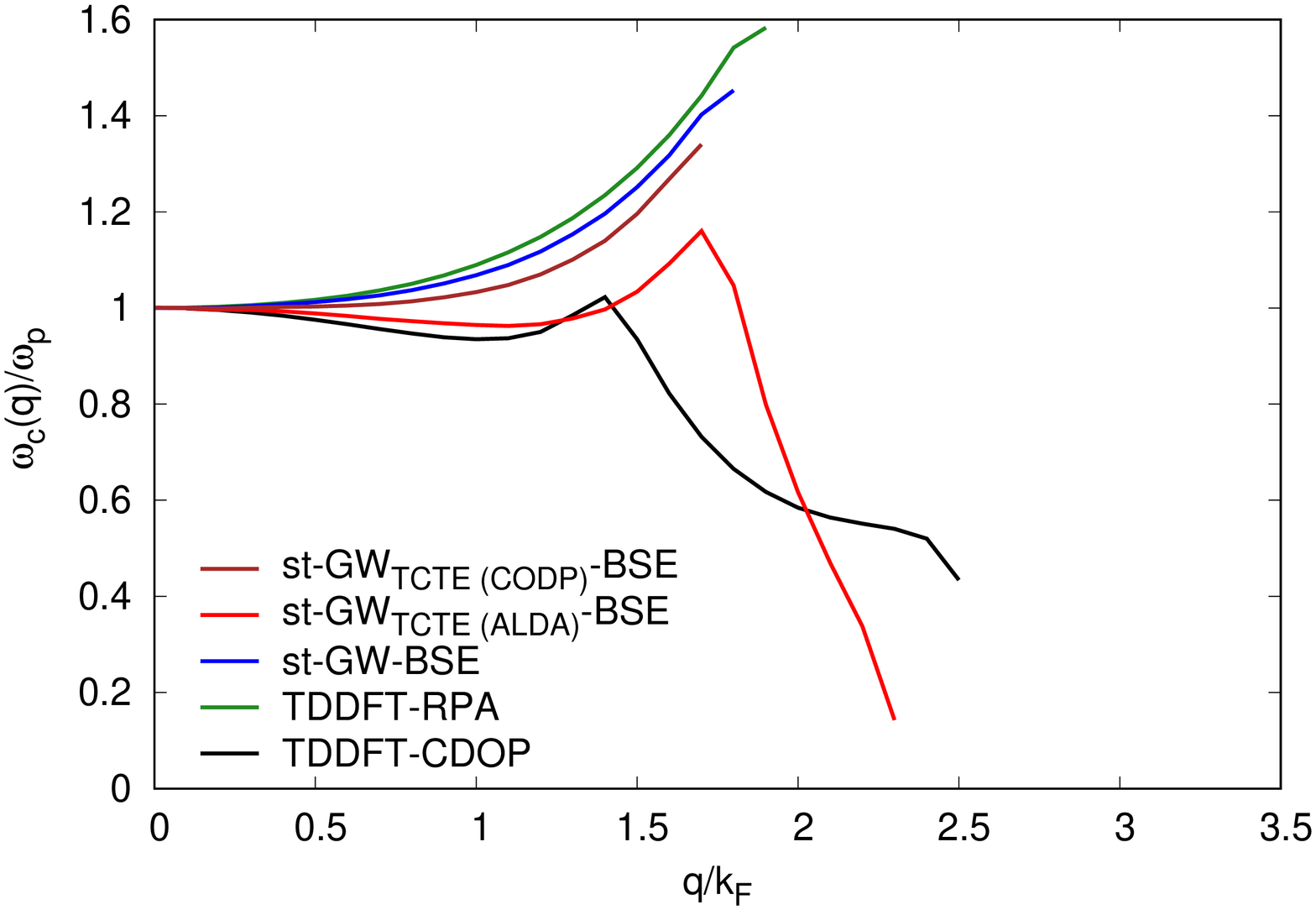}
\caption{Collective modes in the HEG at $r_s=22$ as a function of wavevector.
The lines indicate the position $\omega_c$ of  $\text{Re} \, \epsilon(q,\omega_c(q))=0$.  In green, TDDFT-RPA result.  
In black, TDDFT using the adiabatic CDOP kernel. 
In blue, BSE using a static RPA $W^{\rm in}$ also for the GW self-energy. 
In red, BSE with static TCTE $W^{\rm in}$ calculated with the ALDA kernel, both in the GW self-energy and in the e-h interaction. 
In brown,  BSE with static TCTE $W^{\rm in}$ calculated with the CDOP kernel, both in the GW self-energy and in the e-h interaction. 
}
\label{fig:collective-mode-cdop}
\end{center}
\end{figure}

With respect to the main text, here we 
add results obtained in TDDFT using the CDOP $f_{\rm xc}$ kernel, and in st-GW$_{\rm TCTE}$-BSE with TCTE-CDOP screening.
The st-GW$_{\rm TCTE}$-BSE  yields negative screening with both ALDA and CDOP TCTE vertex corrections (see Fig. \ref{fig:static-screening-cdop}). However, the CDOP $f_{\rm xc}(q)$ kernel is weaker than the ALDA value for $q\neq 0$. Therefore, the negative screening is not pronounced enough in st-GW$_{\rm TCTE (CDOP)}$-BSE, and the dispersion of the collective mode, shown in Fig. \ref{fig:collective-mode-cdop},
remains too close to the RPA plasmon dispersion and does not drop to a low-energy mode at large $q$. Note again that using $f_{\rm xc}$ to simulate the vertex correction is an approximation,
even if the exact $f_{\rm xc}$ is used, and that, moreover, CDOP is accurate only for $\omega=0$. 
Concerning the latter point, the fact that
non-locality and dynamical effects have a tendency to cancel may explain the relative success of the TCTE ALDA vertex correction with respect to the TCTE CDOP one, whereas the fact that both underestimate the effect may be due both to the use of $f_{\rm xc}$ instead of the full vertex correction, and to the approximation of instantaneous screening in the BSE.

\bibliography{heg}